\documentclass[manuscript,screen,authorversion,nonacm,review=false,timestamp=false]{acmart}

\usepackage{fancyhdr}
\AtBeginDocument{%
    \addtolength{\footskip}{2.0\baselineskip}%
    \fancyfoot[L]{\textit{\textbf{Preprint --- do not distribute.}}}%
}
\AtBeginDocument{%
  \providecommand\BibTeX{{%
    \normalfont B\kern-0.5em{\scshape i\kern-0.25em b}\kern-0.8em\TeX}}}

\usepackage{booktabs}
\usepackage{siunitx}

\newcommand{\beginsupplement}{%
        \setcounter{table}{0}
        \renewcommand{\thetable}{S\arabic{table}}%
        \setcounter{figure}{0}
        \renewcommand{\thefigure}{S\arabic{figure}}%
     }

\begin{document}

\title{Phone Sharing and Cash Transfers in Togo: \\
Quantitative Evidence from Mobile Phone Data}

%\title{Mobile Phone Sharing in Togo:\\
%Quantitative Evidence from Nationwide Mobile Phone Data}

\author{Emily L. Aiken}
\authornote{Both authors contributed equally to this research.}
\affiliation{%
  \institution{U.C. Berkeley}
  \country{USA}
}
\email{emilyaiken@berkeley.edu}

\author{Viraj Thakur}
\authornotemark[1]
\affiliation{%
  \institution{U.C. Berkeley}
  \country{USA}
}
\email{viraj.thakur@gmail.com}

\author{Joshua E. Blumenstock}
\affiliation{%
  \institution{U.C. Berkeley}
  \country{USA}
}
\email{jblumenstock@berkeley.edu}

\renewcommand{\shortauthors}{Aiken et al.}

\begin{abstract}
Phone sharing is pervasive in many low- and middle-income countries, affecting how millions of people interact with technology and each other. Yet there is very little quantitative evidence available on the extent or nature of phone sharing in resource-constrained contexts. This paper provides a comprehensive quantitative analysis of phone sharing in Togo, and documents how a large cash transfer program during the COVID-19 pandemic impacted sharing patterns. We analyze mobile phone records from the entire Togolese mobile network to measure the movement of SIM cards between SIM card slots (often on different mobile devices). First, we document the prevalence of sharing in Togo, with 22\% of SIMs and 7\% of SIM slots shared. Second, using administrative data from a government-run cash transfer program, we find that phone sharing is most common among women, young people, and people in rural areas. Finally, we find that the delivery of cash aid via mobile money significantly increases phone sharing among beneficiaries. We discuss the limitations of measuring phone sharing with mobile network data and the implications of our results for future aid programs delivered via mobile money.

%We present a set of quantitative results on phone sharing in Togo using unique identifiers for SIM cards and mobile phones obtained from call detail records. By measuring movement of SIM cards between SIM card slots (frequently in different mobile devices), we find that phone sharing is common in Togo, with 22\% of SIMs and 7\% of SIM slots shared. By matching call detail records to administrative and demographic data from a large cash aid program, we find that phone sharing is most common among women, young people, and people in rural areas. Using data from a randomized controlled trial associated with the aid program, we find that the delivery of cash aid via mobile money significantly increases phone sharing among beneficiaries. We discuss the limitations of measuring phone sharing with call detail records and the implications of our results for future aid programs delivered via mobile money.
\end{abstract}




\maketitle

\section{Introduction}
Over the past two decades, mobile phones have transformed how billions of people communicate and interact. More recently, the rise of mobile money has reshaped everyday financial transactions, including how people send and spend money, apply for loans, and receive humanitarian assistance. In many low- and middle-income countries (LMICs), phone sharing has played an integral role in this digital transformation. Phone sharing is particularly pervasive in rural Africa, where all members of a household or small community sometimes rely on a single primary device \cite{gillwald2005towards,wesolowski2012heterogeneous,blumenstock2010mobile}.

The social practices that surround phone sharing are complex and nuanced. A rich body of geographically diverse qualitative and ethnographic studies highlights how phone sharing is mediated by complex social relationships and obligations \cite{burrell2010evaluating}; can restrict women's empowerment \cite{porter2020mobile}; leads to delicate interpersonal dynamics of reciprocity and payment \cite{burrell2010evaluating, ahmed2017digital}; and helps people create and alter the social spaces around them \cite{steenson2017beyond}. Yet, while there are several qualitative studies of phone sharing dynamics, there is very little quantitative evidence on phone sharing: to our knowledge, only a handful of studies in the past two decades have provided systematic evidence on phone sharing in LMICs, based on sample surveys from Rwanda, Kenya, Namibia, and South Africa \cite{blumenstock2010mobile, wesolowski2012heterogeneous, gillwald2005towards}.

%Mobile phone sharing in resource-constrained contexts has been the subject of sociological investigation for some time \cite{rangaswamy2009personalizing, burrell2010evaluating, dodson2013minding, ahmed2017digital, sey2009exploring, blumenstock2010mobile, wesolowski2012heterogeneous, steenson2017beyond}. A  set of geographically diverse qualitative and ethnographic studies of phone sharing among families, neighborhoods, and communities in low- and middle-income countries (LMICs) have drawn attention to the challenges of adapting mobile technologies for shared use. They have also highlighted how phone sharing patterns are mediated by family and gender hierarchies, and involve nuanced interpersonal dynamics of reciprocity and payment \cite{rangaswamy2009personalizing, burrell2010evaluating, dodson2013minding, ahmed2017digital, sey2009exploring, steenson2017beyond}. While the qualitative literature on phone sharing dynamics is robust, there is a lack of quantitative, representative, and population-scale evidence on phone sharing: only a handful of studies in the past two decades have gathered nationally representative data on patterns of phone sharing in LMICs, with information available on Rwanda, Kenya, Namibia, and South Africa \cite{blumenstock2010mobile, wesolowski2012heterogeneous, gillwald2005towards}.

Gathering more comprehensive evidence on phone sharing is particularly important in the context of the rapid digitization of social protection programs globally. With worldwide mobile phone penetration projected to exceed 70\% in 2025 \cite{gsma}, proponents of aid delivery via mobile money argue that the transparency, traceability, and scalability of mobile aid payments justify a switch from cash to mobile money in the aid sector \cite{davidovic2020beyond, aker2016payment}. Recently, the public health conditions of the COVID-19 pandemic have rendered in-person delivery of aid payments infeasible, leaving mobile money as the de facto standard for remote cash aid delivery in LMICs. Gentillini et al. (2020) \cite{gentilini2020social} estimate that 155 programs in 58 countries delivered aid via digital transfers during the COVID-19 pandemic, prominently including the second wave of the Philippines' Social Amelioration Program (reaching 14 million households),\footnote{https://www.poverty-action.org/printpdf/48011} Pakistan’s Ehsaas emergency cash program (reaching 12 million households),\footnote{https://www.pass.gov.pk/} and Togo’s Novissi program (reaching approximately one million individuals).\footnote{https://novissi.gouv.tg}

While the promise of mobile aid delivery has been discussed at length in the context of the COVID-19 pandemic \cite{davidovic2020beyond, aker2016payment, gronbach2021mobile, bazarbash2020mobile, palacios2020scaling}, it is important to consider how mobile aid delivery interacts with heterogeneous patterns of mobile phone uptake and use, including phone sharing. Existing work on phone sharing suggests that aid delivered via mobile money may not reach the neediest recipients if certain groups (particularly women and the economically vulnerable) are disadvantaged in phone sharing relationships \cite{gillwald2005towards, burrell2010evaluating, wesolowski2012heterogeneous, porter2020mobile}. It is also reasonable to hypothesize that the delivery of cash aid via mobile money may change patterns of phone sharing, possibly increasing human contact when aid is delivered in the context of public health crises like the COVID-19 pandemic.

This paper provides the first population-scale evidence on phone sharing, and the first evidence on the impacts of cash aid delivery via mobile money on phone sharing. We make three main contributions. First, by examining the movement of SIM cards between mobile devices (SIM card slots) as recorded in call detail records (CDR) from Togo, we estimate SIM and device sharing on Togo's entire mobile phone network, finding that 22\% of SIM cards are shared (used by more than one device), and 7\% of devices are shared (used with more than one SIM card). We note that our measure of phone sharing captures only certain types of phone sharing --- specifically, sharing that involves SIM movement between devices --- and therefore likely represents a lower bound on true phone sharing in Togo.  Second, by linking CDR to administrative data, we study heterogeneity in phone sharing patterns by gender, age, and geography. Our findings are largely concurrent with past ethnographic and survey-based work on phone sharing, finding that sharing is highest among women \cite{rangaswamy2009personalizing, burrell2010evaluating, blumenstock2010mobile, wesolowski2012heterogeneous}, young people \cite{wesolowski2012heterogeneous}, and people living in rural areas \cite{sey2009exploring, wesolowski2012heterogeneous}. Finally, we study how phone sharing practices were impacted by a large cash transfer program that was delivered via mobile money. This analysis reveals that the government's mobile money transfers led to a small but significant positive effect on phone sharing, with similar increases across different demographic subgroups. We conclude by discussing the limitations of measuring phone sharing with CDR and the implications of our findings for future aid programs delivered via mobile money.

\subsection{Related Work}

This paper builds on three bodies of literature: studies of phone sharing in LMICs; studies using call detail records to analyze social dynamics, particularly in LMICs; and studies evaluating the impacts of mobile cash transfer programs.

\subsubsection{Phone sharing} 
Several studies have used ethnographic \cite{rangaswamy2009personalizing, burrell2010evaluating, dodson2013minding, ahmed2017digital, sey2009exploring} and survey-based \cite{blumenstock2010mobile, wesolowski2012heterogeneous, gillwald2005towards} approaches to study patterns of phone sharing in a variety of LMIC contexts, ranging from rural Berber communities in Morocco \cite{dodson2013minding} to urban neighborhoods in Bangladesh \cite{ahmed2017digital}. Most geographically relevant to our work in Togo is a study involving qualitative observation and interviews with 22 mobile phone users in neighboring Ghana \cite{sey2009exploring}, which suggests that peer-to-peer phone sharing in Ghana is more common in rural areas than urban ones. It also emphasizes the importance of mobile payphones, which are run by entrepreneurs or agents of the telecommunications companies. People can use mobile payphones to make calls from their phone and/or SIM card for a fee, leading to a set of ``super user" phones and SIM cards that are shared among hundreds or thousands of people. 

Other ethnographic projects in India \cite{rangaswamy2009personalizing}, Uganda \cite{burrell2010evaluating}, Morocco \cite{dodson2013minding}, and Bangladesh \cite{ahmed2017digital} have been more focused on the family and gender dynamics of phone sharing practices. All four studies indicate that women are generally de-prioritized in phone ownership and sharing arrangements within families, with men frequently sharing their personal phones with wives and mothers \cite{rangaswamy2009personalizing, ahmed2017digital, burrell2010evaluating}, or controlling the sharing dynamics of a family phone \cite{rangaswamy2009personalizing, dodson2013minding, ahmed2017digital}. Burrell (2010) \cite{burrell2010evaluating} also analyzes the gender dynamics of phone sharing outside of family units in Uganda, in particular noting that in some cases ``gifting" of phone time between men and women is associated with expectations of reciprocity in the form of money or sexual favors. 

Three studies have used larger-scale survey-based approaches to estimate nationally representative statistics on phone sharing in Africa. Gillwald (2005) \cite{gillwald2005towards} presents data from nationally representative household surveys in ten African countries conducted in 2004 and 2005, although information on phone sharing is only published for four countries. 73.3\% of phone-owning survey respondents in Namibia report sharing their mobile phone with friends at least occasionally; the corresponding statistic is 70\% in Rwanda, 50\% in South Africa, and 48.3\% in Botswana. In data from Namibia, Gillwald (2005) finds that phone sharing is highest among low-income households (for example, 34.9\% of households in the lowest-income quartile regularly share their phone with friends, in comparison to 7.27\% of households in the highest-income quartile). Gillwald (2005) also finds large differences between rural and urban areas in phone sharing, with 61.8\% of households in rural Namibia sharing their phones at least occasionally with friends, in comparison to 10.9\% of households in major towns.  

Wesolowski et al. (2012) \cite{wesolowski2012heterogeneous} analyzes data from a large and nationally representative household survey in Kenya conducted in 2009 (\textit{N} = 32,748), finding that while 85\% of respondents report using a phone, only 44\% own their own phone; phone sharers who do not own their own device are predominantly female (65\%) and the spouses of heads of household (60\%). Wesolowski et al. (2012) also note that phone sharing (relative to ownership) is highest among the poor, young people (under the age of 24), illiterate people, farmers (in comparison to other occupations), and women. Blumenstock and Eagle (2010) \cite{blumenstock2010mobile} conduct a similar exercise with a phone survey of 901 respondents in Rwanda in 2009, reporting that around a third of respondents share their mobile phone, with phone sharing statistically significantly higher among women than men. No significant differences were found in phone sharing by geography or level of education.

\subsubsection{Work with call detail records}
Our work also builds off a set of past studies that have used mobile phone metadata to study social patterns. These studies have used Call Detail Records (CDR, described in Section~\ref{sec:cdr}) to study population density \cite{deville2014dynamic, douglass2015high}, patterns of migration \cite{blumenstock2012inferring, blumenstock2019migration,chi2020general}, friendship formation \cite{onnela2007structure, eagle2009inferring}, social clustering \cite{hidalgo2008dynamics,barnett2016social}, and transportation networks \cite{jarv2012mobile, jahangiri2015applying, graells2018inferring}. Most relevant to our work is the use of CDR to study social and economic dynamics in LMICs, including measuring poverty in Afghanistan \cite{blumenstock2018estimating,aiken2020targeting}, Guatemala \cite{hernandez2017estimating}, Rwanda \cite{blumenstock2015predicting}, and Togo \cite{aiken2021machine}; literacy rates in Senegal \cite{schmid2017constructing}; women's empowerment in Uganda \cite{slavchevska2021can}; and employment shocks in South Asia \cite{sundsoy2017towards}. However, to our knowledge, no research has yet been published using call detail records to study phone sharing.

\subsubsection{The impact of mobile cash transfer programs}
Finally, our work builds on a small literature studying the impact of receiving cash transfers via mobile money. In particular, Aker et al. (2016) \cite{aker2016payment} use a randomized controlled trial to compare the impacts of a cash transfer program delivered via mobile money to the same program delivered in person via cash. In addition to documenting positive impacts on household food security and nutritional status for all beneficiaries, the study documents a 9-16\% increase in food security for households receiving transfers via mobile money in comparison to households receiving transfers via physical cash. The study also notes that women in households that received mobile transfers had greater bargaining power in how the transfer was spent. Blumenstock et al. (2015) \cite{blumenstock2015promises} study a similar RCT through which some employees of a large NGO in Afghanistan were randomly assigned to receive salary payments in mobile money instead of cash. While mobile money payments led to efficiency gains for the employer, there was little evidence of positive benefits to employees.  While these results may be promising for the potential of cash transfer delivery via mobile money, no studies to date have documented how mobile money transfers interact with dynamics of phone sharing within families and communities.

\section{Methods}
This section presents the data and empirical strategies used for our analysis of phone sharing in Togo. Our analysis relies on call detail records (\ref{sec:cdr}) matched to administrative data from the GiveDirectly-Novissi cash aid program in Togo (\ref{sec:admindata}). Parts of our analysis also incorporate data from a large randomized controlled trial associated with the GiveDirectly-Novissi program (\ref{sec:rctdata}). We use regression analysis with these data to estimate (1) demographic differences in phone sharing, and (2) impacts of mobile cash transfers on phone sharing (\ref{sec:merge}, \ref{sec:empiricalstrategy}). We conclude the section by discussing the limitations of measuring phone sharing with CDR (\ref{sec:datalimitations}).\footnote{The code for our analyses is available in the GitHub repository located at https://github.com/emilylaiken/togo-phone-sharing/.}

\subsection{Call Detail Records and Measures of Phone Sharing}
\label{sec:cdr}

Our analysis is based on records of mobile phone transactions in Togo, which were obtained from Togo's two mobile phone operators for the months of October 2020 to January 2021.\footnote{The analysis of these data is governed by a data management plan and research protocol reviewed and approved by U.C. Berkeley's Committee for the Protection of Human Subjects.}
These Call Detail Records (CDR) contain the metadata of each phone call or text message sent from a mobile phone on the Togolese cell network in these five months. We use the following information from both call and text records in our analysis:

\begin{enumerate}
    \item \textit{MSISDN}: Mobile Station International Subscriber Directory Number; the phone number (unique ID) for the SIM card placing the call\footnote{Since phone numbers are personally identifying information, phone numbers are encrypted prior to analysis.}
    \item \textit{IMEI}: International mobile equipment identify; the unique ID for the SIM card slot the MSISDN is in when the call is placed
    \item \textit{Time stamp}: Date and time when the transaction is placed
\end{enumerate}

Note that based on these data, we only observe MSISDN and IMEI for outgoing calls and texts (while our data also record the MSISDN for incoming transactions, they do not record the IMEI for incoming transactions, so incoming transactions are not incorporated in our analysis). Also note that the definition of an IMEI is particularly important: rather than uniquely identifying a physical mobile device, the IMEI uniquely identifies a SIM card slot. This distinction is relevant to countries like Togo, where multi-SIM phones are common.\footnote{While we are not aware of any quantitative studies of the prevalence of multi-SIM phones, based on qualitative field observation they are quite common. In nearby Nigeria, nearly half of all subscribers are estimated to use multi-SIM phones  (https://deviceatlas.com/blog/dual-sim-smartphone-usage).} We discuss the affordances and limitations of this definition further in Section \ref{sec:datalimitations}. 

In total, we observe 1.51 billion transactions (1.06 billion calls and 445.54 million texts) assigned to 5.78 million unique MSISDNs and 9.41 million unique IMEIs between October 1 and January 31, 2021. We remove transactions on seven days that contain anomalous activity (two standard deviations above or below the monthly mean) with respect to the number of active MSISDNs or IMEIs.\footnote{Anomalous days are a mix of days with unusually low activity (likely representing equipment malfunctions causing some mobile phone transactions to be left out of the central databases of the mobile network operators), and days of unusually high activity (for example, there are spikes in activity around Christmas and New Years).}

We construct two measures of phone sharing from these CDR. Our first measure is \textit{SIM sharing}: the extent to which SIMs are moved between devices (SIM card slots). We measure SIM sharing by counting the number of unique IMEIs with which a SIM card (MSISDN) is associated in each month. We consider a SIM to be ``shared" in a given month if it makes transactions with more than one IMEI in that month. Our second measure of phone sharing is \textit{device sharing}: the extent to which devices (SIM card slots, identified by IMEIs) are used with multiple SIMs. We consider a device ``shared" in a given month if it makes transactions associated with more than one MSISDN in that month.

\subsection{The GiveDirectly-Novissi Aid Program and Associated Administrative Data}
\label{sec:admindata}
We complement the call detail records with administrative data from registrations for the GiveDirectly-Novissi aid program in Togo. The GiveDirectly-Novissi program, administered by NGO GiveDirectly and the government of Togo via the government's Novissi social protection platform, provided unconditional cash transfers to hundreds of thousands of Togolese citizens between November 2020 and August 2021. Registration and payment for GiveDirectly-Novissi was arranged entirely via mobile phones: applicants filled out a short USSD (text-based) survey for registration, and eligible applicants received five months of monthly cash transfers via mobile money. Transfers for eligible applicants began immediately following registration. Female beneficiaries received CFA 8,170 (USD 14.40) per month, and male beneficiaries recieved CFA 7,000 (USD 12.34) per month. Aiken et al. (2021) \cite{aiken2021machine} provide more details about the structure and targeting of GiveDirectly-Novissi aid.

\begin{figure}[H]

  \centering
  \includegraphics[width=0.7\linewidth]{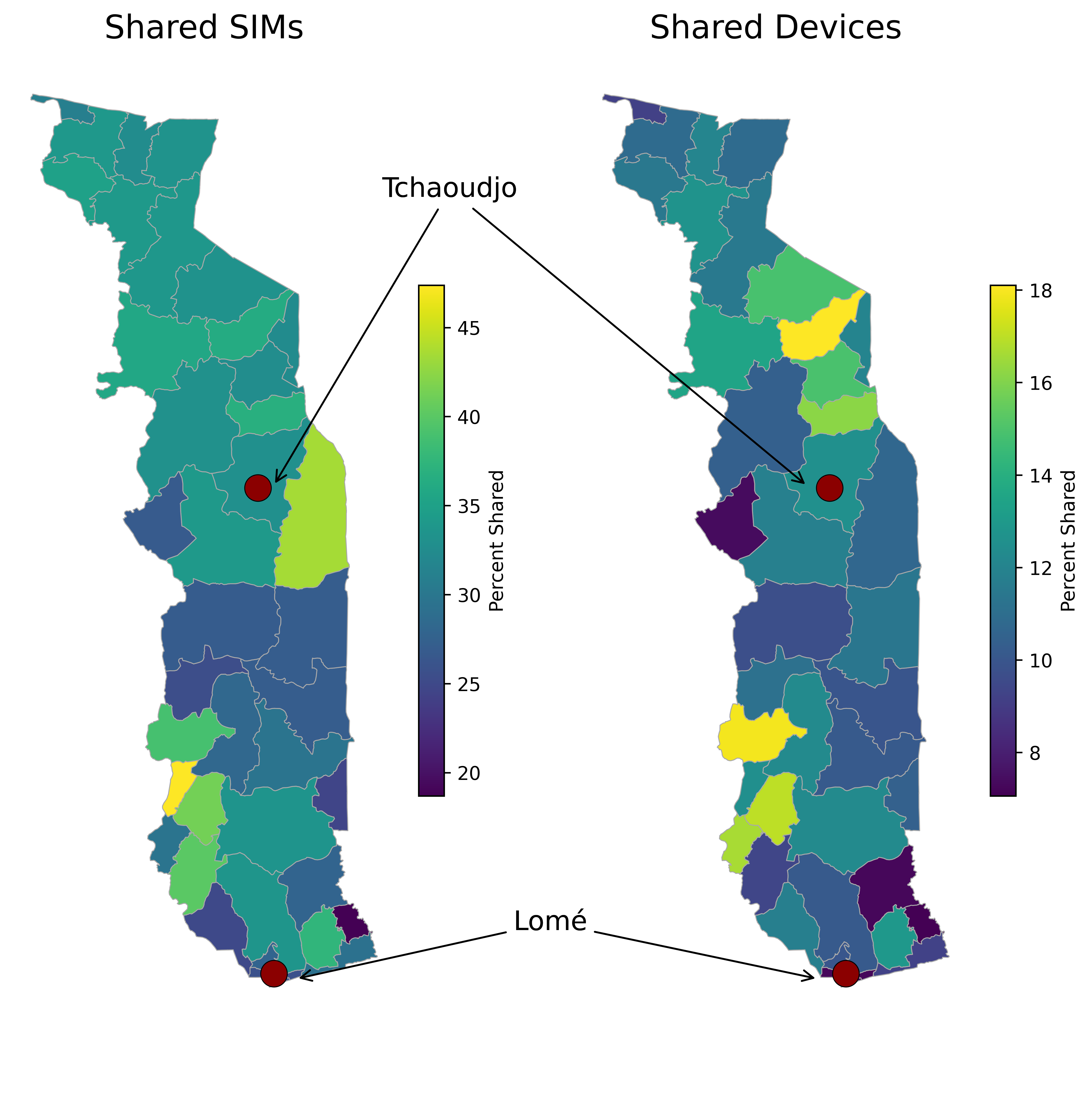}
  
  \caption{\label{figure:geography}SIM sharing (left) and device sharing (right) in October 2020, by prefecture (admin-3 unit, 40 prefectures in Togo).}

\end{figure}

As part of the phone-based registration for GiveDirectly-Novissi aid, applicants were required to provide their voter ID number for unique identification. Aiken et al. (2021) \cite{aiken2021machine} estimate that between 83\% and 98\% of Togolese adults own a voter ID; information for the voter ID was recorded in voter registration drives each year, with the most recent drive occurring in 2020. The administrative information recorded with voter IDs makes it possible to associate the following basic demographic information with each SIM card that registered for GiveDirectly-Novissi:

\begin{enumerate}
    \item \textit{Gender} (male or female)
    \item \textit{Prefecture} of residence (the admin-3 unit; see Figure~\ref{figure:geography})
    \item \textit{Age} 
    \item \textit{Day of registration} to GiveDirectly-Novissi
\end{enumerate}

Note that a single SIM card could only register one voter ID on the USSD registration platform, so each SIM card is uniquely tied to an individual via their voter ID; people with multi-SIM phones could register one person on each SIM. 

In total, 485,977 SIMs registered for GiveDirectly-Novissi between November 2020 and August 2021. We restrict our analysis to 171,530 SIMs registered to GiveDirectly-Novissi in November or December 2020.\footnote{Restricting to registrations in November and December allows us to analyze phone sharing in January among subscribers who received a cash transfer in January, in comparison to a randomly selected treatment group, facilitating our evaluation of the impacts of GD-Novissi transfers on phone sharing (Sections \ref{sec:impacts} and \ref{sec:results_impacts}).} For parts of our analysis --- particularly those focusing on the impacts of cash transfers on phone sharing --- we restrict further to the 49,083 registrations that met the eligibility criteria for GiveDirectly-Novissi benefits.\footnote{Program eligibility was determined by location of residence and the estimated wealth of the registered subscriber \cite{aiken2021machine}. Of the 171,530 GiveDirectly-Novissi registrations in November and December 2020, 149,947 (87.43\%) are in eligible geographies, and 49,083 (34.76\% of registrations in eligible geographies) are below the poverty threshold.} Table \ref{table:demographicstatistics} Column 1 provides a summary of the demographic characteristics of the 171,530 registrations to GiveDirectly-Novissi in November and December 2020. Column 2 provides the demographic breakdown of the 49,083 aid-eligible registrants.

\subsection{Randomized Controlled Trial}
\label{sec:rctdata}

The GiveDirectly-Novissi program was designed as a  randomized controlled trial (RCT), through which some beneficiaries received payments beginning in Fall 2020, and others received payments beginning in Spring 2021. There were two reasons for randomizing the time of eligibility for GiveDirectly-Novissi transfers: first, because the program did not have enough funding to pay all eligible subscribers in Fall 2020, and second, because the program implementers were interested in conducting an impact evaluation of GiveDirectly-Novissi. Our analysis also leverages this randomization to measure the impacts of cash transfers on phone sharing.
 
 Of program registrants in November and December who met the two eligibility criteria (geographic eligibility and poverty-based eligibility), 27,673 (56.38\%) were assigned to the ``treatment group" and were provided with monthly cash payments, with the first payment delivered immediately upon registration (November-December 2020), and the remaining four payments delivered every thirty days. The remaining 21,410 (43.62\%) were assigned to the control group, and received the same total cash aid in lump transfers between May and August 2021. Much of our analysis relies on comparing outcomes between the treatment and control groups, as treatment assignment was entirely at random. Table \ref{table:appendix0} confirms that in October 2020, prior to registration and cash transfer delivery, there was no significant difference between treatment and control groups in terms of demographic information or phone sharing.

\subsection{Merging Datasets}
\label{sec:merge}

\subsubsection{Matching at the SIM level}
\label{sec:msisdnmatching}

For analysis of SIM sharing (whether a SIM is used in multiple devices), we match GiveDirectly-Novissi registrations to phone sharing measures (which are derived from CDR as described in section \ref{sec:cdr}) directly via MSISDN. Both GiveDirectly-Novissi registrations and the SIM sharing measure are uniquely associated with an MSISDN, so matching between administrative data and phone sharing at the SIM level is 1:1.

\subsubsection{Matching at the IMEI level}
\label{sec:imeimatching}

Analysis of device sharing (whether a device is associated with multiple SIM cards) requires linking between GiveDirectly-Novissi registrations and IMEIs. The IMEI from which USSD registration is placed is not recorded in the GiveDirectly-Novissi database, and we do not observe USSD transactions in our CDR, so we are not able to directly match GiveDirectly-Novissi registrations to devices via IMEIs. Instead, we assign each registration to the device used by the registered SIM card on the day closest to program registration. For 56\% of registrants, the ``nearest" device is used on the same day as registration,\footnote{In comparison, only 23\% of registered SIMs make at least one outgoing transaction on the average day in October, suggesting that subscribers are substantially more likely to make a transaction on their day of registration than on a random day.} for 84\% of registrants, it is within one week, and for 97\% of registrants it is within one month. For SIMs that are associated with more than one device on the day of registration (5.50\% of all registrations), we use the device through which the SIM places the most transaction on the closest day. For those with no unique maximum (1.57\%), we take one of the maximum devices at random. 

Note that this matching strategy means that matching between registrations and measures of device sharing is 1:N --- one device can be associated with multiple registrations, if it is the closest device to the date of registration for multiple SIMs. 3.34\% of devices in our dataset are linked to more than one SIM; in our analysis of device sharing they appear once in the dataset for each registration they are linked to. In Tables \ref{table:appendix1}, \ref{table:appendix2}, and \ref{table:appendix3}, we evaluate the sensitivity of our main results to this 1:N matching strategy, finding that all results remain significant at the 0.05 level if we (1) include devices associated with multiple registered SIMs only once in the dataset, picking the associated registration at random, and (2) drop devices associated with multiple registered SIMs altogether. 

\subsection{Empirical Strategy}
\label{sec:empiricalstrategy}

We rely on regression analysis throughout for evaluating (1) differences in phone sharing by demographic characteristics, and (2) impacts of treatment on phone sharing.

\subsubsection{Heterogeneity in phone sharing}
\label{sec:heterogenetiy}
To estimate differences in patterns of phone sharing by different demographic subgroups (gender, age, and location), we use the dataset of GiveDirectly-Novissi registrations in November and December matched to mobile phone data from October 2020, before the aid program was launched on November 1, 2020. Matching to CDR from October allows us to assess heterogeneity in phone sharing before behavior was impacted by GiveDirectly-Novissi. At the SIM level, this dataset contains 122,173 observations (after removing SIMs that do not place at least one outgoing transaction in October). At the device level, this dataset contains 105,443 observations (after removing devices that do not place at least one outgoing transaction in October). 

Our analysis quantifies the magnitude and significance of the differences in these two measures of phone sharing between (1) men and women, (2) urban areas (the prefectures of Lomé Commune, Agoe-Nyive, and Golfe --- which make up the capital city of Lomé --- and the prefecture of Sokodé which contains the second largest city of Tchaoudjo) and rural areas (the remaining 36 prefectures), and (3) four age groups (age 18-30, 30-40, 40-50, and 50+).\footnote{In order to obtain a voter ID, applicants had to be at least age 18.} Throughout, we employ a basic specification (regressing phone sharing only on the relevant demographic characteristic), and a specification that controls for the number of transactions placed by each SIM or device (as our measure of phone sharing may be confounded by the number of transactions placed, see Section \ref{sec:datalimitations}). These regression specifications are shown in Equation \ref{eq:1}, where $y_i$ is an indicator for whether observation $i$ is a shared SIM (or device)\footnote{We prefer these binary measures of phone sharing to, for example, the average counts of devices per SIM and SIMs per deivce that are included in Table \ref{table:summarystatistics}, as the average count can be susceptible to a few ``super-user" devices that are associated with many SIMs. For example, among 105,443 devices matched to GiveDirectly-Novissi registration data in October 2020, 165 devices (0.16\%) are associated with more than 1,000 SIMs in the month of October, skewing the average measures high. It is possible that these ``super-user" SIMs correspond to the payphones identified in previous work in Ghana \cite{sey2009exploring} that are rented out for a price to people who do not own phones.}, $x_i$ is the demographic variable of interest, $n_i$ is the number of transactions placed by observation $i$ (included only in the specification controlling for number of transactions), and $\epsilon_i$  is an idiosyncratic error term.

\begin{equation}
\label{eq:1}
    y_i = \beta_0 + \beta_1x_i + \beta_2n_i + \epsilon_i
\end{equation}
\vspace{1pt}

\subsubsection{Impact of mobile cash transfers on phone sharing}
\label{sec:impacts}
To estimate the impact of cash transfers delivered via mobile money on phone sharing, we restrict our dataset to subscribers enrolled in the RCT (i.e. SIMs and devices that are associated with registrations that met the geographic and poverty requirements for GiveDirectly-Novissi, and were assigned to either treatment or control groups). 

We first employ a basic specification that uses phone sharing data only from the month of January 2021, by which time all subscribers in the treatment group had received at least one cash transfer. At the SIM level, this dataset contains 43,323 observations (after removing SIMs that do not place at least one outgoing transaction in January). At the device level, this dataset contains 30,483 observations (after removing devices that do not place at least one outgoing transaction in January). Treatment was assigned entirely at random for the GiveDirectly-Novissi RCT, so a simple regression on treatment is sufficient to estimate the treatment effect of cash transfers on phone sharing. As in the earlier demographic analyses, we experiment with controlling for the number of transactions. We also explore the addition of demographic controls. These regression specifications are shown in Equation \ref{eq:2}, which includes an indicator $T_i$ for whether individual $i$ is in the treatment group. 

\begin{equation}
\label{eq:2}
    y_i = \beta_0 + \beta_1T_i + \beta_2n_i + \beta_3x_i + \epsilon_i
\end{equation}
\vspace{1pt}

We also employ a difference-in-differences specification in order to better isolate changes in phone sharing over time (for a given SIM or device) caused by the cash transfer. In the difference-in-differences specification, each observation represents a subscriber-month ($N$ = 173,654 at the SIM level, $N$ = 147,419 at the device level), using phone sharing data from October, November, December, and January. We regress monthly phone sharing $y_{it}$, an indicator for whether observation $i$ is shared in month $t$, on $T_{it}$, an indicator for whether observation $i$ received a cash transfer in month $t$, while controlling for MSISDN fixed effects $\mu_i$ and month fixed effects $\pi_t$, as well as $n_{it}$, the number of transactions placed by observation $i$ in month $t$. We cluster standard errors at the SIM level for SIM sharing and the device level for device sharing.

\begin{equation}
\label{eq:3}
    y_{it} = \beta_0 + \beta_1T_{it} + \beta_2\pi_t + \beta_3\mu_i + \beta_4n_{it} + \epsilon_{it}
\end{equation}
\vspace{1pt}

We return to the basic regression specification and data from January 2020 to test for heterogeneous treatment effects by gender (men vs. women) and age (under 30, 30-40, 40-50, or over 50).\footnote{We cannot test for heterogeneous treatment effects by geography (urban vs. rural), as only rural areas were eligible for GiveDirectly-Novissi aid (see Table \ref{table:demographicstatistics}.)} Specifically, to measure heterogeneous treatment effects by gender, we regress SIM and device sharing on treatment, number of transactions, a binary indicator for being female, and the indicator for being female crossed with treatment. We employ an analogous specification for testing for heterogeneous treatment effects by age group. These specifications are shown in Equation \ref{eq:4}, where $x_i$ is the demographic information of interest (gender or age group). 

\begin{equation}
\label{eq:4}
    y_i = \beta_0 + \beta_1T_i + \beta_2n_i + \beta_3x_i + \beta_4T_i \times x_i + \epsilon_i
\end{equation}

\subsection{Data Limitations}
\label{sec:limitations}

Before proceeding with the results of these analyses, we note several limitations of the data and methods employed. These include:

\begin{enumerate}
    \item \textit{Limits to observation of phone sharing in CDR:} The CDR only allow us to observe phone sharing that involves a SIM card moving between devices (SIM card slots). We therefore cannot observe other common types of phone sharing, for instance when a single device and SIM card is shared between multiple people (without switching SIM cards). In this sense, our measure of phone sharing is a substantial underestimate of ``true" phone sharing in Togo. We likewise cannot track when subscribers change phones or SIM cards.\footnote{By using the phone number (MSISDN) as the unique identifier for SIM cards, we can keep track of subscribers who switch SIM cards but transfer their phone number to the new SIM. For this reason we use MSISDNs as our unique SIM identifier rather than IMSI (international mobile subscriber identity numbers), which cannot be transferred between SIM cards.}
    
    \item \textit{Limits to IMEIs as device identifiers:} IMEIs, our ``device identifiers", uniquely identify SIM card slots rather than mobile devices. Our analysis would therefore count the movement of a single SIM card between SIM card slots in the same device as a type of phone sharing, so SIM sharing may be overestimated where SIMs are moved within a physical device. Device sharing may be underestimated where multiple SIMs are used with a single device, but in different SIM card slots.
    
    \item \textit{Phone sharing is confounded by transaction count}: Our measures of phone sharing --- whether a SIM is used in multiple SIM card slots and whether a SIM card slot is associated with multiple SIMs --- may be confounded by the number of transactions placed by an SIM or device (SIM card slot). Phone sharing can only be observed for subscribers placing at least one transaction; SIMs or devices placing only one transaction in a given month will be assigned by definition as ``not shared." To address this concern, we provide specifications that control for the number of transactions placed by an SIM or device throughout.
    
    \item \textit{Imperfect matching between GiveDirectly-Novissi registrations and SIM cards}: GiveDirectly-Novissi registrations are linked to SIM cards according to the voter ID registered for GiveDirectly-Novissi with each SIM card. However, it is not required that the owner of a phone register their voter ID on that phone, and in some cases there may not be a designated owner for each SIM card or device. For example, based on qualitative field observation of GiveDirectly-Novissi registrations, it is common that in households with only a single phone, the device and SIM cards inside it are owned by the man, but he registers his wife's voter ID for GiveDirectly-Novissi (women receive more aid than men, which incentivizes this behavior). Given these patterns, the matching between SIMs and Novissi registration data reflects only associations rather than ownership.
    
    \item \textit{Imperfect matching of GiveDirectly-Novissi registrations and devices}: In addition to the issues with linking SIMs to GiveDirectly-Novissi registrations described in (4) above, there is some further noise in matching devices to GiveDirectly-Novissi registrations. As described in Section \ref{sec:imeimatching}, the device (IMEI) of registration is not recorded in Novissi registrations, and USSD transactions for registration are not recorded in CDR. We therefore match registrations to the ``nearest" device (IMEI) associated with the SIM card of registration, but in some cases it is likely that this matching does not reflect the device that the registration was truly performed on.
    
    \item \textit{Limits to data representativity}: Our dataset of CDR matched to GiveDirectly-Novissi registrations is not nationally representative of Togo, nor representative of the parts of Togo in which GiveDirectly-Novissi aid was distributed. There is evidence of considerable self-selection in who registers for GiveDirectly-Novissi aid \cite{aiken2021machine}, and structural elements of the program preclude certain populations from registering and therefore being included in our analysis (for example, those without a voter ID or without access to a phone could not register). Our analysis of CDR matched to registration data therefore can only be interpreted as representative of the population registering for GiveDirectly-Novissi; likewise our analysis of treatment effects is only representative of the population eligible for GiveDirectly-Novissi aid (where the eligible population is on average poorer and more vulnerable than the applicant population \cite{aiken2021machine}).
    
\end{enumerate}
\label{sec:datalimitations}

\section{Results}
\subsection{Summary Statistics on Phone Sharing}
\label{sec:results_summary}

Table \ref{table:summarystatistics} presents summary statistics on SIM and device sharing in Togo using CDR from October 2020, prior to the launch of GiveDirectly-Novissi in November 2020. We present results for three groups: (1) all SIM cards and devices on Togo's two mobile phone networks, (2) all SIM cards and devices that registered for the GiveDirectly-Novissi cash transfer program in November and December 2020, and (3) all SIM cards and devices that met the eligibility criteria for  the GiveDirectly-Novissi RCT (see Section \ref{sec:rctdata}).
\footnote{There are fewer observations in column 2 of Table~\ref{table:summarystatistics} than in Table~\ref{table:demographicstatistics} because we can only measure phone sharing in the month of October for SIM cards and devices that placed at least one transaction during the month that is recorded in our CDR; not all registered subscribers make a transaction in every month.}
We find substantial evidence of SIM and device sharing: 22.38\% of devices are shared, and 7.44\% of devices are shared. There is higher phone sharing among those registered for GiveDirectly-Novissi (33.13\% shared SIMs and 11.98\% shared devices), perhaps because advertisement for the program was targeted towards rural areas. We note, however, that across groups the median and modal SIM card and device are not shared. 

\subsection{Heterogeneity in Phone Sharing}
\label{sec:results_heterogeneity}

Table \ref{table:demographics} evaluates differences in phone sharing among demographic groups using data from October 2020, prior to the launch of the GiveDirectly-Novissi aid program in November 2020. We examine heterogeneity by (1) gender (men vs. women), (2) age group (18-30, 30-40, 40-50, and 50+), and (3) geography (urban vs. rural, with urban defined as prefectures --- admin-3 units --- containing the two major cities in Togo, and rural defined as all other prefectures). We find that after controlling for the number of transactions, which can confound our measures of phone sharing, SIM cards associated with women are not statistically significantly more likely to be shared, but devices associated with women are 1.84 percentage points (17.10\%) more likely to be shared than those associated with men (\textit{p} < 0.001). We find that SIMs and devices associated with people age 18-30 are 6.91-8.47 points (26.09-3.98\%) and 1.99-3.86 points (17.20-39.75\%) more likely to be shared than those associated with people over thirty (\textit{p} < 0.001), respectively. Finally, we find that SIMs  in rural areas are 10.03 points (53.27\%) more likely to be shared than SIMs in urban areas, and devices in rural areas are 3.16 points (36.92\%) more likely to be shared than devices in urban areas (\textit{p} < 0.001).

Figure \ref{figure:geography} further illustrates heterogeneous phone sharing by geography, by mapping the degree of SIM and device sharing by prefecture (admin-3 units). While there are heterogeneous patterns of variation throughout the country, in general SIM and device sharing are higher in the Northern rural parts of the country in comparison with the more urban areas in the South.

\subsection{Cash Transfer Impacts on Phone Sharing}
\label{sec:results_impacts}

To measure the impacts of receiving cash transfers via mobile money on phone sharing, we restrict our dataset to SIMs and devices enrolled in the GiveDirectly-Novissi RCT and use CDR from January 2021. The treatment group for the RCT ($N$=27,673) received cash transfers every month for five months after registration; since we restrict our analysis to subscribers registered in November or December 2020, all treated subscribers received a cash transfer in January 2021. The control group ($N$=21,410) began receiving cash transfers in May 2021, so no control subscribers received cash in January 2021. 

 The first two columns of Table \ref{table:impacts} present the results of regressing phone sharing in January on treatment status. We find that after controlling for number of transactions, treated SIM cards are 1.92 percentage points (8.16\%) more likely to be shared in January than control SIM cards (\textit{p} < 0.0001). Treated devices are 0.94 points (12.32\%) more likely to be shared in January than control devices (\textit{p} < 0.01). We further find that the magnitude and significance of these effects are robust to the inclusion of a variety of demographic controls (Column 4).   

\begin{figure}[H]

  \centering
  \includegraphics[width=0.6\linewidth]{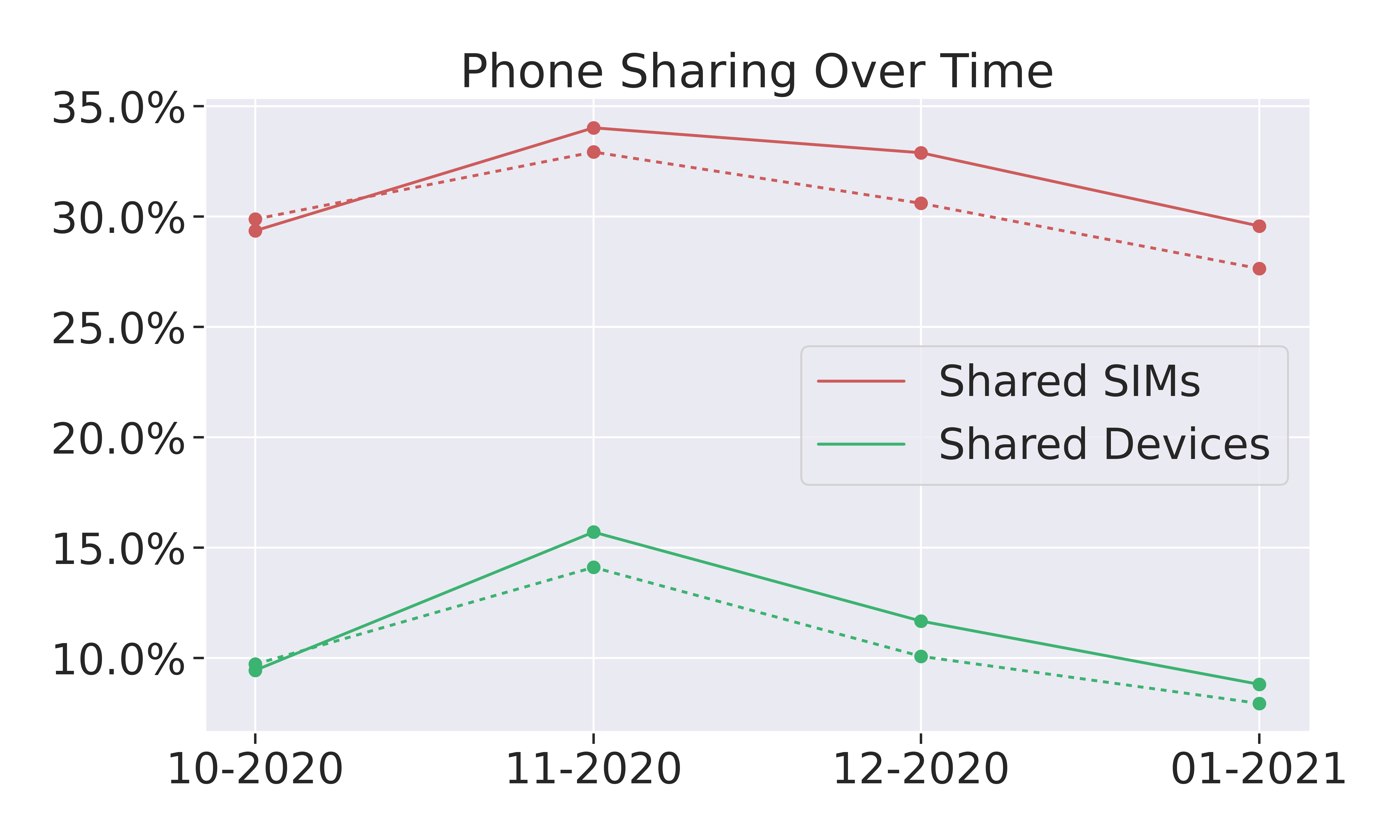}
  
  \caption{\label{figure:temporality}SIM sharing (red) and device sharing (green) from October to January 2020, separately for the treatment group (solid line) and control group (dashed line).}

\end{figure}

To measure changes in phone sharing over time caused by cash transfers, we use a difference-in-differences specification with CDR from October, November, December, and January.\footnote{We do not observe CDR after January so our analysis is restricted to these four months, however, treated subscribers continued to receive payments through April.} Figure \ref{figure:temporality} provides visual motivation for the differences-in-differences specification, plotting the percentage of SIMs and devices shared by month, separately for treatment and control groups. While treatment and control groups are nearly indistinguishable for both measures of phone sharing in October, before the aid program began, the treatment group has markedly more phone sharing in November, December, and January, when cash transfers were distributed to some or all treatment subscribers.

Observations for the differences-in-differences specification are SIM-months (or device-months), and subscriber $i$ is ``treated" in month $t$ if subscriber $i$ received a cash transfer at anytime in month $t$; the differences-in-differences specification also includes month and subscriber fixed effects. Table \ref{table:diffindiff} presents the results of the differences-in-differences specification. As in the earlier set of results we observe positive and significant impacts on phone sharing from treatment (treatment effects of 2.86 percentage points, or 11.19\%, for SIM sharing and 2.10 percentage points, or 21.63\%, for device sharing, \textit{p} < 0.001). 

\subsubsection{Heterogeneous treatment effects}
We return to the basic regression specification to test for heterogeneous treatment effects by gender (men vs. women) and age (under 30, 30-40, 40-50, or over 50). Table \ref{table:heteroimpacts} shows no statistically significant differences in impacts of cash transfers on phone sharing by gender or age group.

\section{Discussion}
This section expands on two particularly notable aspects of our results: the concordance of our results with past work, in spite of differences in measurement methods and study locations; and the implications of our results for future aid programs delivered via mobile money.

\subsection{Concordance with past work}

In Section \ref{sec:results_heterogeneity} we find that both SIM and device sharing are most common among women, young people, and people in rural areas. Specifically, focusing on device sharing, phones associated with women are 17.10\% more likely to be shared than phones associated with men; phones associated with people under 30 are 17.20-39.75\% more likely to be shared than phones associated with people over age 30, and phones in rural areas are 36.92\% more likely to be shared than phones in urban areas. 

The gender dynamic has been studied extensively, both quantitatively and qualitatively, in past work on phone sharing. Our finding that women use shared phones more than men is consistent with Wesolowski et al. (2012) \cite{wesolowski2012heterogeneous}, which finds that nearly twice as many women as men in Kenya report using a shared phone rather than a privately owned one (65\% of phone sharers are women, and 35\% are men). It is also consistent with Blumenstock et al. \cite{blumenstock2010mobile}, in which 39\% of female respondents to a phone survey in Rwanda report sharing the phone on which they answered the survey, in comparison to 25\% of men. Our work is also consistent with a set of qualitative and ethnographic studies that have documented higher phone sharing among women than men in Bangladesh, India, and Uganda \cite{rangaswamy2009personalizing, burrell2010evaluating, ahmed2017digital}. 

Differences in phone sharing by age groups has been studied less thoroughly in past work. Our finding that young people (those under 30) share phones more than older people (over 30) is, however, consistent with results from Kenya in Wesolowski et al. \cite{wesolowski2012heterogeneous}, which finds that 51\% of survey respondents ages 16-17 report using a shared phone (in comparison to 17\% phone ownership in the same group), 41\% of respondents aged 18-24 report using a shared phone (in comparison to 43\% phone ownership), and 31-34\% of respondents aged 25-59 report using a shared a phone (in comparison to 41-53\% ownership). It is worth noting that Wesolowski et al. find that phone sharing increases again after age 59, with 39-42\% of respondents aged 60 and above reporting using a shared a phone (in comparison with 23-29\% phone ownership). We observe a similar pattern, with phone sharing highest among those under 30, followed by those over 50, then those ages 40-50, and finally lowest among those aged 30-40. It is possible that high levels of phone sharing among young people reflect that young people cannot afford a personal phone, or that age hierarchies in families or communities cause parents to share phones with children rather than purchasing phones for young people (as observed by Ahmed at al. (2017) in Bangladesh \cite{ahmed2017digital}). 

Finally, our finding on differences in phone sharing by geography is also consistent with Gillwald (2005) \cite{gillwald2005towards} in Namibia and Wesolowski et al. (2012) \cite{wesolowski2012heterogeneous} in Kenya, both of which find that phone sharing is more common in rural areas than urban ones. It is also consistent with Sey (2009) \cite{sey2009exploring}, which suggests that phone sharing in Ghana is higher in rural areas than in cities, though conclusions are limited by a small sample size (\textit{N} = 22). The large difference we identify in phone sharing between urban and rural areas in Togo is not consistent with Blumenstock and Eagle (2010), which finds no statistically significant difference between urban and rural phone sharing in Rwanda. 

The similarity in results between our work and past work on phone sharing, particularly Wesolowski et al (2012), is especially notable considering that the method used for measuring phone sharing in our paper is quite different from methods employed in past studies. In particular, our CDR-based measure of phone sharing only captures sharing that involves the movement of SIM cards between devices, uses the SIM card slot as the unique device ID rather than an ID for a physical mobile phone, and is subject to several sources of imperfect matching between phone sharing measures produced from CDR and demographic data (Section \ref{sec:limitations}). The similarities between our results and past work suggest that patterns of phone sharing can be quantitatively measured with CDR despite these limitations. 

\subsection{Implications for policy}
Our results, which document the impact of cash transfers on phone sharing and highlight demographic heterogeneity in phone sharing, have implications for future policies on mobile delivery of aid payments. To recap, we find that the delivery of cash aid via mobile money significantly increases phone sharing, with subscribers in the randomly assigned treatment group in the GiveDirectly-Novissi program using a shared SIM card 8.16-11.19\% more often than subscribers in the control group, and using a shared device 12.32-21.63\% more frequently than subscribers in the control group. While these increases in phone sharing are unlikely to negate all the health and economic benefits of receiving cash transfers via a mobile phone  \cite{davidovic2020beyond, aker2016payment}, the increase in person-to-person contact caused by mobile aid delivery may be a concern for public health officials in the context of a pandemic.

Moreover, our results highlight that it is important to consider phone sharing dynamics in the design of mobile aid programs. If people who rely on shared phones are less able to receive benefits than people with personal devices, programs that deliver aid via mobile money risk disadvantaging groups that are frequent sharers of mobile phones, including women, young people, and people in rural areas. One option to address this issue is to allow any number of people to receive cash transfers on a mobile phone --- though this strategy may be open to fraud (malicious actors registering other people on their phone and keeping the payments). A middle ground is to limit registrations on a single SIM card to a single individual, and provide financing for phone sharers to purchase their own SIM card.\footnote{SIM cards are typically much cheaper than mobile devices. The GiveDirectly-Novissi program limited registrations on a single SIM card to one individual.} It is worth noting, however, that even if policies are designed to ensure that phone sharers can register for aid programs, phone sharers are at higher risk of having their aid payments taken by other users of the same phone. This issue highlights the need for mobile-based user interfaces that allow for private digital zones in the context of shared use of mobile financial systems \cite{ahmed2017digital, ahmed2019everyone}, as well as the importance of foregrounding the needs of vulnerable groups in the design of aid delivery systems.

\bibliographystyle{ACM-Reference-Format}
\bibliography{sample-base}

\clearpage
\section*{Figures and Tables}

% \usepackage{booktabs}
\begin{table}[]
\begin{tabular}{@{}lcccc@{}}
\toprule
                                              & \multicolumn{2}{c}{\textbf{Matched to Administrative Data}} & \multicolumn{2}{c}{\textbf{Enrolled in RCT}} \\
                                              & Number                          & Proportion                        & Number                    & Proportion                   \\ \midrule
\textit{Panel A : Gender}                     &                              &                              &                      &                       \\
Female                                        & 95,396                       & 55.61\%                      & 24,694               & 50.31\%               \\
Male                                          & 76,134                       & 44.39\%                      & 24,389               & 49.69\%               \\ \\
\textit{Panel B: Age Group}                   &                              &                              &                      &                       \\
18-30                                         & 67,424                       & 39.30\%                      & 16,081               & 32.76\%               \\
30-40                                         & 43,606                       & 25.42\%                      & 14,770               & 30.09\%               \\
40-50                                         & 29,415                       & 17.15\%                      & 9,428                & 19.21\%               \\
50+                                           & 31,095                       & 18.13\%                      & 8,804                & 17.94\%               \\ \\
\textit{Panel C: Geography (urban vs. rural)} &                              &                              &                      &                       \\
Rural                                         & 167,436                      & 97.61\%                      & 49,083               & 100.00\%               \\
Urban                                         & 4,094                        & 2.39\%                       & 0                    & 0.00\%                \\ \\
\textit{N} (Total)                                            & 171,530                     &                   & 49,083               &          \\ \bottomrule
\end{tabular}

\caption{\label{table:demographicstatistics}Summary statistics on demographic breakdown for (1) GiveDirectly-Novissi registrations in November and December (left), which we use for our analysis of heterogeneity in phone sharing among demographic groups, and (2) GiveDirectly-Novissi beneficiaries enrolled in the RCT (right), which we use for our analysis of impacts of cash transfers on phone sharing.}
\end{table}

\begin{table}[H]
\begin{tabular}{@{}lccc@{}}
\toprule
                     & \textbf{Entire network} & \textbf{Matched to administrative data} & \textbf{Enrolled in RCT} \\ \midrule
\multicolumn{4}{l}{\textit{Panel A: SIM sharing (Devices per SIM)}}                             \\
Mean                 & 1.35 (0.85)    & 1.60 (1.15)                    & 1.50 (1.00)     \\
Median               & 1.00           & 1.00                           & 1.00            \\
Mode                 & 1.00           & 1.00                           & 1.00            \\
\% of SIMs shared & 22.38\%        & 33.13\%                        & 29.58\%         \\
\textit{N}                    & 4,508,881      & 122,173                       & 41,738          \\ \\

\multicolumn{4}{l}{\textit{Panel B: Device sharing (SIMs per device)}}                           \\
Mean                 & 1.09 (1.86)    & 4.24 (63.19)                   & 3.88 (60.67)    \\
Median               & 1.00           & 1.00                           & 1.00            \\
Mode                 & 1.00           & 1.00                           & 1.00            \\
\% of devices shared   & 7.44\%         & 11.98\%                        & 9.56\%          \\
\textit{N}                    & 5,600,581      & 105,443                     & 35,500          \\ \bottomrule
\end{tabular}

\caption{\label{table:summarystatistics}Summary statistics on phone sharing in October 2020 for (1) the entire mobile phone network (left), (2) observations matched to administrative data from GiveDirectly-Novissi (center), and (3) observations enrolled in the RCT (right). For the measures of mean devices per SIM and mean SIMs per device, standard deviations are shown in parentheses.}

\end{table}

% \usepackage{booktabs}
\begin{table}[H]
\begin{tabular}{@{}lccccccc@{}}
\toprule
                              & \multicolumn{2}{c}{\textbf{Gender}}    & \multicolumn{2}{c}{\textbf{Age}}       & \multicolumn{2}{c}{\textbf{Geography}} & \textbf{Joint} \\
                              & Basic & Control & Basic & Control & Basic & Control & \textbf{}      \\ \midrule
\textit{Panel A: SIM sharing} &                &                       &                &                       &                &                       &                \\
Constant                      & 0.3419***      & 0.2875***             & 0.3832***      & 0.3340***             & 0.2733***      & 0.1883***             & 0.2309***      \\
Number of mobile transactions                   &                & 0.0008***             &                & 0.0008***             &                & 0.0008***             & 0.0008***      \\
Female                        & -0.0205***     & -0.0020               &                &                       &                &                       & 0.0027         \\
Age: 30-40                    &                &                       & -0.0919***     & -0.0847***            &                &                       & -0.0851***     \\
Age: 40-50                    &                &                       & -0.0839***     & -0.0732***            &                &                       & -0.0742***     \\
Age: 50+                      &                &                       & -0.0814***     & -0.0691***            &                &                       & -0.0708***     \\
Rural                         &                &                       &                &                       & 0.0597***      & 0.1003***             & 0.1043***      \\
\textit{$R^2$}                   & 0.0005         & 0.0254                & 0.0082         & 0.0319                & 0.0005         & 0.0267                & 0.0333         \\
\textit{N}                    & 122,173        & 122,173               & 122,173        & 122,173               & 122,173        & 122,173               & 122,173        \\ \\

\multicolumn{8}{l}{\textit{Panel B: Device sharing}}                                                                                                                      \\
Constant                      & 0.1104***      & 0.1076***             & 0.1380***      & 0.1356***             & 0.0893***      & 0.0856***             & 0.0924***      \\
Number of mobile transactions                   &                & 0.0000***             &                & 0.0000***             &                & 0.0000***             & 0.0000***      \\
Female                        & 0.0182***      & 0.0184***             &                &                       &                &                       & 0.0200***      \\
Age: 30-40                    &                &                       & -0.0383***     & -0.0386***            &                &                       & -0.0396***     \\
Age: 40-50                    &                &                       & -0.0284***     & -0.0291***            &                &                       & -0.0307***     \\
Age: 50+                      &                &                       & -0.0196***     & -0.0199***            &                &                       & -0.0216***     \\
Rural                         &                &                       &                &                       & 0.0316***      & 0.0316***             & -0.0347***     \\
\textit{$R^2$}                   & 0.0008         & 0.01769                & 0.0024         & 0.0185                & 0.0003         & 0.0164                & 0.0199         \\
\textit{N}                    & 105,443        & 105,443                & 105,443         & 105,443                & 105,443         & 105,443                & 105,443         \\ \bottomrule
\end{tabular}

\caption{\label{table:demographics}Heterogeneity in patterns of phone sharing by (1) gender (male vs. female), (2) age group (18-30, 30-40, 40-50, and 50+, and (3) geography (urban vs. rural). For each outcome variable, we present two specifications: one regressing on just the demogrpahic variable(s) (left), and one regressing on the demographic variable(s) and controlling for the number of transactions associated with each SIM or device (right). In the far right column, we include a joint specification with all demographic information included. Statistical significance of regression coefficients are shown with stars: * $\rightarrow$ $\textit{p}$ < 0.05, ** $\rightarrow$ $\textit{p}$ < 0.01, *** $\rightarrow$ $\textit{p}$ < 0.001.}

\end{table}

% \usepackage{booktabs}
\begin{table}[H]
\begin{tabular}{@{}lccc@{}}
\toprule
                                 & \textbf{Basic} & \textbf{Transactions Control} & \textbf{Demographic Controls} \\ \midrule
\textit{Panel A: SIM sharing}    &                &                               &                               \\
Constant                         & 0.2764***      & 0.2352***                     & 0.2831***                     \\
Treatment (received cash transfer)  & 0.0193***      & 0.0192***                     & 0.0196***                     \\
Number of mobile transactions                    &                & 0.0008***                     & 0.0008***                     \\
Female                           &                &                               & 0.0140***                     \\
Age: 30-40                       &                &                               & -0.0749***                    \\
Age: 40-50                       &                &                               & -0.0856***                    \\
Age: 50+                         &                &                               & -0.0846***                    \\
\textit{$R^2$}                      & 0.0004         & 0.0194                        & 0.0268                        \\
\textit{N}                       & 43,323         & 43,323                      & 43,323                       \\ \\
\textit{Panel B: Device sharing} &                &                               &                               \\
Constant                         & 0.0794***      & 0.0763***                     & 0.0882***                     \\
Treatment (received cash transfer)  & 0.0087**      & 0.0094**                     & 0.0096**                     \\
Number of mobile transactions                   &                & 0.0000***                     & 0.0000***                     \\
Female                           &                &                               & 0.0063*                       \\
Age: 30-40                       &                &                               & -0.0183***                    \\
Age: 40-50                       &                &                               & -0.0236***                    \\
Age: 50+                         &                &                               & -0.0255***                    \\
\textit{$R^2$}                      & 0.0002         & 0.0244                       & 0.0267                        \\
\textit{N}                       & 30,483         & 30,483                        & 30,483                         \\ \bottomrule
\end{tabular}

\caption{\label{table:impacts} Impacts of cash transfers delivered via mobile money on SIM sharing (panel A) and device sharing (panel B) in January 2020. We employ three regression specifications: the \textit{basic specification} (left) which regresses phone sharing outcomes only on treatment, the \textit{transactions controls specification} (middle) which controls for number of outgoing transactions by a SIM or device in January 2020, and the \textit{demographic controls specification} which controls for the number of transactions alongside a set of demographic controls obtained from administrative data. Statistical significance of regression coefficients are shown with stars: * $\rightarrow$ $\textit{p}$ < 0.05, ** $\rightarrow$ $\textit{p}$ < 0.01, *** $\rightarrow$ $\textit{p}$ < 0.001.}

\end{table}

% \usepackage{booktabs}
\begin{table}[H]
\begin{tabular}{@{}lcc@{}}
\toprule
                                 & \textbf{Basic} & \textbf{Transactions Control} \\ \midrule
\textit{Panel A: SIM sharing}    &                &                               \\
Constant                         & 0.2962***      & 0.2556***                     \\
Treatment x Month                & 0.0305***      & 0.0286***                     \\
Number of mobile transactions    &                & 0.0009***                     \\
Month: November                  & 0.0218**       & 0.0230***                     \\
Month: December                  & 0.0050        & 0.0019                        \\
Month: January                   & -0.0254***     & -0.0299***                    \\
\textit{$R^2$}                   & 0.0031         & 0.0095                       \\
\textit{N}                       & 173,654         & 173,654                        \\ \\
\textit{Panel B: Device sharing} &                &                               \\
Constant                         & 0.0975***      & 0.0971***                     \\
Treatment x Month                & 0.0210***      & 0.0210***                     \\
Number of mobile transactions    &                & 0.0000***                     \\
Month: November                  & 0.0391***      & 0.0391***                     \\
Month: December                  & 0.0000         & 0.0001                        \\
Month: January                   & -0.0241***     & -0.0239***                    \\
\textit{$R^2$}                   & 0.0106         & 0.0105                       \\
\textit{N}                       & 147,419         & 147,419                      \\ \bottomrule
\end{tabular}

\caption{\label{table:diffindiff}Difference-in-differences specification for estimating the impacts of cash transfers delivered via mobile money on SIM sharing (Panel A) and device sharing (Panel B). We employ two differences-in-differences specifications: a basic one (left), and one controlling for the number of outward transactions an SIM or device places in each month (right). Statistical significance of regression coefficients are shown with stars: * $\rightarrow$ $\textit{p}$ < 0.05, ** $\rightarrow$ $\textit{p}$ < 0.01, *** $\rightarrow$ $\textit{p}$ < 0.001. Standard errors are clustered at the MSISDN level for Panel A and the IMEI level for Panel B.}

\end{table}

% \usepackage{booktabs}
\begin{table}[H]
\begin{tabular}{@{}lcccc@{}}
\toprule
                       & \multicolumn{2}{c}{\textit{Panel A: SIM sharing}} & \multicolumn{2}{c}{\textit{Panel B: Device sharing}} \\
                       & \textbf{Gender}             & \textbf{Age}                 & \textbf{Gender}               & \textbf{Age}                  \\ \midrule
Constant               & 0.2274***          & 0.2910***           & 0.0735***            & 0.0910***            \\
Treatment (received cash transfer) & 0.0135**         & 0.0204*           & 0.0072               & 0.0109               \\
Number of mobile transactions           & 0.0008***          & 0.0008***           & 0.0000***            & 0.0000***            \\
Female                 & 0.0140*            &                     & 0.0057               &                      \\
Age: 30-40             &                    & -0.0743***          &                      & -0.0171**           \\
Age: 40-50             &                    & -0.0918***          &                      & -0.0188**          \\
Age: 50+               &                    & -0.0810***          &                      & -0.0313***           \\
Female x Treatment     & 0.0115             &                     & 0.0046               &                      \\
Age: 30-40 x Treatment &                    & -0.0025             &                      & -0.0029              \\
Age: 40-50 x Treatment &                    & 0.0079              &                      & -0.0097              \\
Age: 50+ x Treatment   &                    & -0.0088             &                      & -0.0089              \\
$R^2$                     & 0.0199            & 0.0266              & 0.0246               & 0.0260               \\
N                      & 43,323             & 43,323              & 30,483               & 30,483               \\ 
\bottomrule
\end{tabular}

\caption{\label{table:heteroimpacts} Heterogeneous impacts of mobile cash transfers on SIM sharing (Panel A) and device sharing (Panel B) by gender (left) and age group (right) in January 2020. For each specification we control for the number of mobile transactions associated with the SIM or device in January 2020. Statistical significance of regression coefficients are shown with stars: * $\rightarrow$ $\textit{p}$ < 0.05, ** $\rightarrow$ $\textit{p}$ < 0.01, *** $\rightarrow$ $\textit{p}$ < 0.001.}
\end{table}

\clearpage
\beginsupplement
\appendix
\label{appendix}
\section*{Supplementary Tables}

% \usepackage{booktabs}
\begin{table}[H]
\begin{tabular}{@{}lccc@{}}
\toprule
                       & \textbf{Treatment} & \textbf{Control} & \textbf{Difference} \\ \midrule
\% Women               & 0.50 (0.50)        & 0.51 (0.50)      & -0.01               \\
Age                    & 38.01 (13.49)      & 37.91 (13.53)    & 0.10                \\
Shared SIM             & 0.29 (0.46)        & 0.30 (0.46)      & -0.01               \\
Shared device          & 0.09 (0.29)        & 0.10 (0.30)      & 0.00              \\
Number of mobile transactions & 45.47 (70.02)      & 45.93 (71.32)    & -0.46               \\
\textit{N}             & 27,673             & 21,410           & 49,083              \\ \bottomrule
\end{tabular}

\caption{\label{table:appendix0}Pre-program differences between treatment and control groups for the GiveDirectly-Novissi program, using Novissi registration data along with mobile phone data from October 2020. Statistical significance of the difference in means is determined with a t-test, and significance is shown with stars: * $\rightarrow$ $\textit{p}$ < 0.05, ** $\rightarrow$ $\textit{p}$ < 0.01, *** $\rightarrow$ $\textit{p}$ < 0.001 (no differences are significant). }.

\end{table}

\begin{table}[H]
\begin{tabular}{lcc}
\hline
                   & \textbf{Matched to Administrative Data} & \textbf{Enrolled in RCT} \\ \hline
\multicolumn{3}{l}{\textit{Panel A: 1:N Device Matching}}                                      \\
Mean               & 4.24 (63.19)                            & 3.92 (60.67)             \\
Median             & 1.00                                    & 1.00                     \\
Mode               & 1.00                                    & 1.00                     \\
\% of devices shared & 11.98\%                                 & 9.56\%                   \\
\textit{N}                  & 105,443                                 & 35,500                   \\ \\
\multicolumn{3}{l}{\textit{Panel B: 1:1 Device Matching}}                                      \\
Mean               & 1.33 (13.80)                            & 1.22 (8.28)              \\
Median             & 1.00                                    & 1.00                     \\
Mode               & 1.00                                    & 1.00                     \\
\% of devices shared & 11.24\%                                 & 8.92\%                   \\
\textit{N}                  & 101,475                                 & 34,332                   \\ \\
\multicolumn{3}{l}{\textit{Panel C: Uniquely Matched Devices Only}}                       \\
Mean               & 1.15 (1.88)                             & 1.11 (0.84)              \\
Median             & 1.00                                    & 1.00                     \\
Mode               & 1.00                                    & 1.00                     \\
\% of devices shared & 10.76\%                                 & 8.54\%                   \\
\textit{N}                  & 98,152                                  & 33,281            \\
\bottomrule
\end{tabular}

\caption{\label{table:appendix1}Sensitivity of the summary statistics on device sharing in October 2020 in Table \ref{table:summarystatistics} to the strategy for matching devices to administrative data. Panel A reproduces the results shown in Panel B of Table \ref{table:summarystatistics}, using the 1:N matching strategy that we use throughout the main paper. Panel B reproduces the same statistics, using a 1:1 matching strategy where for devices that match to more than one phone number (and therefore more than one registration entry), one phone number is selected at random for inclusion in the dataset. Panel C again reproduces the same statistics, this time dropping any registrations associated with devices that match more than one phone number}.  

\end{table}

\begin{table}[H]
\begin{tabular}{lccccccc}
\hline
                                    & \multicolumn{2}{c}{\textbf{Gender}} & \multicolumn{2}{c}{\textbf{Age}} & \multicolumn{2}{c}{\textbf{Geography}} & \textbf{Joint} \\
                                    & Basic            & Control          & Basic           & Control        & Basic              & Control           &                \\ \hline
\textit{Panel A: 1:N Device Matching} &                  &                  &                 &                &                    &                   &                \\
Constant                      & 0.1104***      & 0.1076***             & 0.1380***      & 0.1356***             & 0.0893***      & 0.0856***             & 0.0924***      \\
Number of mobile transactions                   &                & 0.0000***             &                & 0.0000***             &                & 0.0000***             & 0.0000***      \\
Female                        & 0.0182***      & 0.0184***             &                &                       &                &                       & 0.0200***      \\
Age: 30-40                    &                &                       & -0.0383***     & -0.0386***            &                &                       & -0.0396***     \\
Age: 40-50                    &                &                       & -0.0284***     & -0.0291***            &                &                       & -0.0307***     \\
Age: 50+                      &                &                       & -0.0196***     & -0.0199***            &                &                       & -0.0216***     \\
Rural                         &                &                       &                &                       & 0.0316***      & 0.0316***             & -0.0347***     \\
\textit{$R^2$}                   & 0.0008         & 0.01769                & 0.0024         & 0.0185                & 0.0003         & 0.0164                & 0.0199         \\
\textit{N}                    & 105,443        & 105,443                & 105,443         & 105,443                & 105,443         & 105,443                & 105,443         \\ \\
\multicolumn{8}{l}{\textit{Panel B: 1:1 Device Matching}}                                                                                                                \\
Constant                            & 0.1037***        & 0.1020***        & 0.1308***       & 0.1293***      & 0.0864***          & 0.0839***         & 0.0910***      \\
Number of mobile transactions                      &                  & 0.0000***        &                 & 0.0000***      &                    & 0.0000***         & 0.0000***      \\
Female                              & 0.0170***        & 0.0175***        &                 &                &                    &                   & 0.0191***      \\
Age: 30-40                          &                  &                  & -0.0379***      & -0.0379***     &                    &                   & -0.0389***     \\
Age: 40-50                          &                  &                  & -0.0294***      & -0.0292***     &                    &                   & -0.0307***     \\
Age: 50+                            &                  &                  & -0.0200***      & -0.197***      &                    &                   & -0.0212***     \\
Rural                               &                  &                  &                 &                & 0.0269***          & 0.0280***         & -0.0301***     \\
\textit{$R^2$}                         & 0.0007           & 0.0029           & 0.0025          & 0.0047         & 0.0002             & 0.0024            & 0.0059         \\
\textit{N}                          & 101,478          & 101,475          & 101,475        & 101,475        & 101,475            & 101,475           & 101,475       \\ \\
\multicolumn{8}{l}{\textit{Panel C: Uniquely Matched Devices Only}}                                                                                                      \\
Constant                            & 0.0992***        & 0.0918***        & 0.1261***       & 0.1196***      & 0.0861***          & 0.0737***         & 0.0796***      \\
Number of mobile transactions                        &                  & 0.0001***        &                 & 0.0001***      &                    & 0.0001***         & 0.0001***      \\
Female                              & 0.0162***        & 0.0187***        &                 &                &                    &                   & 0.0202***      \\
Age: 30-40                          &                  &                  & -0.0377***      & -0.0369***     &                    &                   & -0.0379***     \\
Age: 40-50                          &                  &                  & -0.0296***      & -0.0284***     &                    &                   & -0.0298***     \\
Age: 50+                            &                  &                  & -0.0209***      & -0.0194***     &                    &                   & -0.0209***     \\
Rural                               &                  &                  &                 &                & 0.0222***          & 0.0287***         & 0.0311***      \\
\textit{$R^2$}                         & 0.0007           & 0.0036           & 0.0026          & 0.0051         & 0.0003             & 0.0029            & 0.0065         \\
\textit{N}                          & 98,152           & 98,152           & 98,152          & 98,152         & 98,152             & 98,152            & 98,152        \\
\bottomrule
\end{tabular}

\caption{\label{table:appendix2}Sensitivity of the results on heterogeneity in patterns of device sharing by gender (left), age(middle), and geography (right) in Table \ref{table:demographics} to the strategy for matching devices to administrative data. Panel A reproduces the results shown in Panel B of Table \ref{table:demographics}, using the 1:N matching strategy that we use throughout the main paper. Panel B and Panel C reproduce the same regression results, using the 1:1 matching strategy and restricting to only uniquely matched devices, respectively.}

\end{table}

% \usepackage{booktabs}
\begin{table}[]
\begin{tabular}{@{}lcc@{}}
\toprule
                                    & \textbf{Basic} & \textbf{Transactions Control} \\ \midrule
\textit{Panel A: 1:N Device Matching} &                &                               \\
Constant                            & 0.0794***      & 0.0763***                     \\
Treatment (received cash transfer)  & 0.0087**      & 0.0094**                     \\
Number of mobile transactions       &                & 0.0000***                     \\
\textit{$R^2$}                      & 0.0002         & 0.0244                        \\
\textit{N}                          & 30,483         & 30,483                        \\ \\
\multicolumn{3}{l}{\textit{Panel B: 1:1 Device Matching}}                              \\
Constant                            & 0.0726***      & 0.0693***                     \\
Treatment (received cash transfer)  & 0.0081**       & 0.0086**                      \\
Number of mobile transactions       &                & 0.0001***                     \\
\textit{$R^2$}                      & 0.0002         & 0.0040                        \\
\textit{N}                          & 29,472         & 29,472                        \\ \\
\multicolumn{3}{l}{\textit{Panel C: Uniquely Matched Devices Only}}                    \\
Constant                            & 0.0685***      & 0.0607***                     \\
Treatment (received cash transfer)  & 0.0072*        & 0.0074*                       \\
Number of mobile transactions       &                & 0.0002***                     \\
\textit{$R^2$}                      & 0.0002         & 0.0036                        \\
\textit{N}                          & 28,615         & 28,615                        \\ \bottomrule
\end{tabular}

\caption{\label{table:appendix3}Sensitivity of the results on the impacts of cash transfers on device sharing in Columns 1 and 2 of Table \ref{table:impacts} to the strategy for matching devices to administrative data. Panel A reproduces the results shown in Panel B of Table \ref{table:impacts}, using the 1:N matching strategy that we use throughout the main paper. Panel B and Panel C reproduce the same regression results, using the 1:1 matching strategy and restricting to only uniquely matched devices, respectively.}

\end{table}

\end{document}